\documentclass[sigconf]{acmart}

\setcopyright{none}
\renewcommand\footnotetextcopyrightpermission[1]{}
\makeatletter
\let\@authorsaddresses\@empty
\makeatother
\AtBeginDocument{%
  }

\usepackage{tabularx}
\usepackage{enumitem}
\usepackage{multirow}
\usepackage{pifont}
\usepackage{siunitx}
\usepackage{dcolumn}
\usepackage[breakable]{tcolorbox}
\usepackage{makecell}

\begin{document}

\author{Synthia Qia Wang}
\affiliation{%
   \institution{Johns Hopkins University}
   \city{Baltimore}
   \state{MD}
   \country{USA}}

\author{Ranya Sharma}
\affiliation{%
   \institution{University of Chicago}
   \city{Chicago}
   \state{IL}
   \country{USA}}

\author{Christine Lee}
\affiliation{%
   \institution{University of Chicago}
   \city{Chicago}
   \state{IL}
   \country{USA}}

\author{Sarah Keedy}
\affiliation{%
   \institution{University of Chicago}
   \city{Chicago}
   \state{IL}
   \country{USA}}

\author{Marshini Chetty}
\affiliation{
  \institution{University of Chicago}
  \city{Chicago}
  \state{IL}
  \country{USA}
}

\author{Nick Feamster}
\affiliation{
  \institution{University of Chicago}
  \city{Chicago}
  \state{IL}
  \country{USA}
}

\title{Measuring Personal Information Exposure in Mental Health-Related Conversations with LLMs}

\begin{abstract}
Large language models (LLMs) are increasingly used for informal mental health support, yet the privacy implications of conversational interaction remain poorly understood. Prior LLM privacy research has focused largely on model-side risks such as memorization and adversarial extraction, leaving personal information inferred from ordinary conversation understudied. We introduce a profile-recovery framework that measures personal information exposure by comparing user profiles with attributes inferred from conversation alone. We generate 240 synthetic multi-turn mental health conversations from profiles sampled from a nationally representative health survey and validate them against real counseling transcripts, finding comparable overall disclosure volume. Across 21 attributes, 33--41\% of each profile is recoverable, with 76.5\% first-inference and 94.3\% any-of-3 accuracy. 239 of 240 conversations expose at least one sensitive mental health attribute, and 45.0\% expose all six. Our findings reveal a gap between disclosure and inferability, motivating HCI approaches to make inference risks legible and support privacy-aware interaction design.
\end{abstract}
\maketitle
\section{Introduction}

Large language models (LLMs) are increasingly used in everyday contexts that involve emotionally sensitive and highly personal discussions. General-purpose systems such as ChatGPT are frequently used for informal mental health support, including emotional reflection, coping strategies, and guidance during distress~\cite{bucher2025s,callahan2026real,yang2026exploring,american2025use}. Although these systems were not originally designed as therapeutic tools, their accessibility, conversational fluency, and perceived neutrality make them appealing to users seeking immediate and stigma-free support. For individuals who face barriers to traditional mental health care such as cost, limited availability of providers, or geographic constraints, LLMs provide an easily accessible alternative for discussing personal challenges and mental health concerns~\cite{bucher2025s,iftikhar2025llm,firth2017efficacy}. As a result, LLMs are increasingly embedded in contexts where conversations involve deeply personal experiences, emotions, and life circumstances. This growing reach has attracted broader attention to potential harms, including concerns that LLM systems may be repurposed to extract personal information from public-facing interactions~\cite{brewster2023chatgpt}, raising important questions about how privacy is affected by what users share and what systems can infer from those interactions.

A defining feature of mental health support interactions is self-disclosure. Sharing personal information about emotions, experiences, and social contexts is central to therapeutic dialogue and relationship-building. Prior work shows that conversational agents can foster meaningful interactions and encourage users to disclose sensitive information, even in the absence of human providers~\cite{darcy2021evidence,lee2020hear}. Design features such as perceived empathy, reciprocity, and human-like communication cues can further increase users' willingness to share personal details~\cite{lee2020hear,cohn2024believing,ischen2019privacy}. With the emergence of highly fluent LLMs capable of sustaining long-form dialogue, these interactions increasingly resemble therapeutic conversations rather than traditional human--computer exchanges. Such interactions may therefore create substantial opportunities for users to disclose personal information, particularly when conversational systems foster trust and psychological safety~\cite{bi2023create,shim2024unveiling,choi2025privacy,peng2026ai}.

Privacy risks, however, extend beyond what users explicitly disclose. Research shows that linguistic patterns and contextual signals in text can enable the inference of sensitive personal attributes, including demographic characteristics, mental health indicators, and behavioral traits~\cite{staab2024beyond,mireshghallah2025position}. In mental health contexts specifically, models have demonstrated the ability to infer conditions such as depression, stress, and suicidal ideation from textual data~\cite{hua2025scoping,wang2024scale,bzdok2024data,xu2024utilizing}. Thus, a conversation can expose information that a user never explicitly states including seemingly ordinary descriptions of relationships, work, living situations, or emotional experiences may collectively provide evidence about latent characteristics. This creates a fundamental distinction between \emph{disclosure}, what users communicate, and \emph{exposure}, what a system can recover from that communication. In multi-turn conversations, these signals accumulate over time, potentially making personal information increasingly inferable even when users do not intend to reveal it, and making it particularly difficult for users to anticipate or prevent~\cite{wang2025beyond,shao2024privacylens}.

Despite increasing attention to privacy risks in LLM systems, existing research has primarily focused on vulnerabilities associated with model behavior and training data. Prior work has demonstrated risks related to memorization, membership inference, and prompt-based extraction of sensitive training data~\cite{shokri2017membership,carlini2021extracting,crothers2023machine,das2025security}. While these studies reveal important attack surfaces, they largely analyze adversarial interactions or model-side leakage rather than privacy risks arising from ordinary user conversations. Research on attribute inference shows that LLMs can derive sensitive personal information from user-generated text even when it is not explicitly stated~\cite{staab2024beyond,dou-etal-2024-reducing,mireshghallah2024trust}, and real-world evidence suggests that a substantial portion of user queries already contain identifying information~\cite{mireshghallah2024trust}. However, existing evaluations largely rely on static datasets or short-form text, providing limited insight into how personal information accumulates across realistic multi-turn conversations. More importantly, we lack a systematic way to quantify how much of a user's underlying personal profile can be recovered from a conversation alone. This gap is particularly consequential in mental health contexts, where conversations naturally involve sensitive experiences and interconnected aspects of users' lives.

This gap motivates three research questions:
\begin{itemize}
    \item \textbf{RQ1:} Can synthetic conversations generated through LLM role-playing serve as a valid proxy for real mental health dialogues, reproducing realistic patterns of self-disclosure?
    \item \textbf{RQ2:} How much personal information can be recovered from a single mental health conversation with an LLM, and how accurately can an external inference model recover that information from conversational content alone?
    \item \textbf{RQ3:} What factors moderate personal information exposure, including conversation length, distress level, and attribute type?
\end{itemize}

Using real conversations for this purpose would require collecting sensitive demographic and clinical information from real individuals and could introduce additional privacy risks through data sharing and publication. We therefore use synthetic data to construct controlled conversations from detailed participant profiles without exposing real individuals. This approach allows us to systematically vary attribute distributions, measure recovery against known ground truth, and reproduce experiments across conditions at scale. Although synthetic interactions may not capture every nuance of real dialogue, prior work has shown that carefully designed synthetic datasets can reproduce key properties of real-world conversational data~\cite{synthpai,de2025introducing,xue2026mindchat,xu2025mentalchat16k}. We consequently treat synthetic conversations not as a replacement for real counseling data, but as a controlled environment for measuring privacy exposure that would be difficult to study safely at comparable scale with real users.

To address our research questions, we introduce a profile-recovery framework for measuring personal information exposure in conversational systems. We sample participant profiles from nationally representative health survey data, generate realistic multi-turn mental health conversations through LLM-based role-playing, and use an independent inference model to recover attributes from conversational content alone. Our conversation-generation protocol was developed with a licensed psychotherapist on our research team, who provided clinical expertise in designing and assessing realistic mental health interactions. Comparing inferred attributes with the known profile provides a controlled measure of how much personal information is recoverable from an interaction and how accurately it can be inferred. We validate our synthetic corpus against real counseling transcripts across linguistic, emotional, and psycholinguistic dimensions, and systematically analyze disclosure and inference patterns across 240 conversations and 21 personal attributes.

Our findings demonstrate substantial and widespread personal information exposure from a single mental health conversation. Across 21 attributes, 33--41\% of each profile is recoverable, with 76.5\% first-inference accuracy and 94.3\% any-of-3 accuracy across inferred fields. 239 of 240 conversations expose at least one sensitive mental health attribute, and 45.0\% expose all six of our most privacy-sensitive attributes. Critically, exposure is not concentrated in longer conversations or among the most distressed users, suggesting that inference risk can arise even during brief interactions and across levels of distress. These findings extend the study of LLM privacy beyond memorization and adversarial extraction to the inference of personal information from ordinary, good-faith interaction. More broadly, our profile-recovery framework provides a controlled approach for measuring the gap between what users disclose and what conversational systems can infer. This gap has direct implications for HCI, that privacy mechanisms for conversational systems should help users understand not only what they share, but also what may be inferred from how they communicate.
\section{Background}
General-purpose LLMs such as ChatGPT are increasingly used for mental health support, despite not being designed as clinical systems~\cite{bucher2025s,callahan2026real,yang2026exploring, american2025use}. Users turn to these systems for emotional support, self-reflection, and crisis management, particularly when cost, geography, stigma, or limited availability create barriers to traditional mental health care~\cite{bucher2025s,iftikhar2025llm, firth2017efficacy}. Unlike many conventional human--computer interactions, these encounters can involve open-ended, emotionally intimate conversations in which users share personal experiences and sensitive information. The conversational fluency and apparent empathy of LLMs may further encourage such disclosure, making the privacy implications of these interactions especially important in mental health contexts~\cite{yang2026exploring,callahan2026real}.

\subsection{LLMs for Mental Health and Self-Disclosure}
Self-disclosure is particularly important in mental health interactions. Prior work shows that conversational agents (CAs) can foster therapeutic relationships and encourage users to disclose personal information, with perceived reciprocity and agent self-disclosure further increasing willingness to share~\cite{darcy2021evidence, lee2020hear}. Studies of LLM-based and other conversational systems similarly find that anthropomorphic, persuasive, and human-like characteristics can increase trust and disclosure of sensitive information~\cite{cohn2024believing, ischen2019privacy, stock2023tell, zhan2025malicious, zhang2024s}. Human-like cues and empathetic responses can amplify disclosure even among users who are otherwise privacy-conscious~\cite{bi2023create, shim2024unveiling, choi2025privacy}, while multimodal cues such as voice and visual features can further strengthen trust and psychological safety~\cite{peng2026ai}.

These interactional benefits also introduce risks. LLM-based mental health systems may over-validate harmful beliefs or create artificial rapport that encourages increasingly sensitive disclosures~\cite{iftikhar2025llm}. Personification and contextual framing can also alter model behavior; for example, providing models with user-specific attributes such as mental health disclosures can change safety responses, while remaining vulnerable to adversarial prompting~\cite{yildirim2026differential}. Other work has raised concerns about over-reliance, emotional attachment, and reduced human oversight in AI-mediated mental health interactions~\cite{asman2025responsible,tavory2024regulating,bolpagni2025development,peter2025benefits}, as well as privacy, data misuse, and unintended exposure of sensitive information~\cite{hua2025scoping,ni2025scoping,uptegraft2025elastic}.

Users may also have limited understanding of how LLMs process, store, or infer information from their inputs~\cite{kimbel2024security, DSIT2024PublicAttitudes, wang2025beyond}, which can contribute to unintentional disclosure. Real-world deployments suggest that users often approach general-purpose LLMs as sources of immediate support or guidance rather than replacements for human care, while emphasizing the importance of safety, transparency, and ethical data use~\cite{callahan2026real}. At the same time, users may incorrectly assume that sensitive conversations receive protections comparable to therapist-patient disclosures and interpret human-like emotional responses as evidence that these systems carry corresponding responsibilities toward users~\cite{kwesi2025exploring}.

\subsection{Personal Information Inference from Conversational Data}
Disclosure, however, captures only part of the privacy risk in conversational systems. Prior work demonstrates that seemingly benign conversational data can reveal latent attributes such as demographic characteristics, mental states, and behavioral patterns~\cite{staab2024beyond, mireshghallah2025position}. In mental health contexts, LLMs have been shown to infer sensitive attributes such as depression, stress, and suicidal ideation from textual data~\cite{hua2025scoping, wang2024scale,bzdok2024data,xu2024utilizing}. More recent work demonstrates that LLMs can identify language biomarkers associated with isolation and hopelessness, further illustrating the extent to which naturalistic dialogue contains psychologically meaningful signals~\cite{yang2026exploring, rasool2025emotion}.

In clinical and applied settings, LLMs have also demonstrated potential for structuring and interpreting health-related information, including classification tasks in which performance may vary across categories~\cite{cardamone2025classifying}. Similarly, LLMs have been used to annotate and extract explainable signals from social media data for diagnostic purposes~\cite{belcastro2025detecting}. More broadly, studies of clinical and social media data indicate that structured and unstructured conversations can be mined for highly sensitive signals, sometimes without explicit user intent~\cite{pavlopoulos2024overview,andibanbang2025ai}. These findings suggest that privacy exposure cannot be characterized solely by examining what users explicitly disclose, as an interaction may also make additional personal attributes recoverable through downstream inference.

\subsection{Synthetic Data for Dialogue and Privacy Evaluation}
Synthetic data provides a way to study conversational systems in domains where real-world data are scarce, sensitive, or difficult to annotate. In mental health research, synthetic datasets have been used to simulate therapy dialogues, augment training data, and evaluate model behavior under controlled conditions~\cite{adhikary2024exploring,ghanadian2024socially,de2025introducing, yin2025mdd, xu2025mentalchat16k}. Systems such as MindChat use synthetic multi-turn counseling corpora generated through role-playing frameworks to improve model capabilities while also investigating privacy properties such as susceptibility to membership inference attacks~\cite{xue2026mindchat}. More broadly, studies show that LLM-generated data can achieve strong performance in tasks including sentiment analysis, suicide risk detection, and counseling simulation, sometimes rivaling real-world datasets~\cite{ghanadian2024socially,hasan2025early, yin2025mdd,lozoya2025synthetic}. Synthetic datasets have likewise been proposed as a resource for privacy research when confidentiality constraints make real-world data difficult to use~\cite{staab2024beyond, synthpai}.

However, existing uses of synthetic data primarily focus on generating realistic dialogue, training models, or evaluating model behavior. They do not directly address how much personal information a conversation can expose or how accurately that information can subsequently be reconstructed when the user's underlying attributes are known. In sensitive domains such as mental health, this controlled setting provides an opportunity to study privacy exposure while retaining access to the ground truth needed to evaluate what can actually be recovered from an interaction.

\subsection{Research Gap}
Much of the LLM privacy literature has focused on model-side risks, including memorization, membership inference, and extraction of sensitive training data~\cite{shokri2017membership, carlini2021extracting, crothers2023machine, das2025security}. Studies have demonstrated that sensitive information can be reproduced or exposed through prompt injection, jailbreaking, and direct querying~\cite{perez2022ignore, liu2023jailbreaking, lukas2023analyzing, li-etal-2023-multi-step}. These approaches are important for understanding model vulnerabilities, but they address a different privacy question from whether users' own conversational disclosures can be used to reconstruct personal information.

Attribute-inference research moves closer to this question by showing that sensitive personal information can be derived from user-generated text, even when such information is not explicitly stated~\cite{staab2024beyond, dou-etal-2024-reducing, mireshghallah2024trust}. Real-world evidence further suggests that a substantial portion of user queries already contain identifying information~\cite{mireshghallah2024trust}, while studies of privacy-preserving input strategies show that users may struggle to anticipate and prevent implicit inference from their text~\cite{shao2024privacylens, synthpai, wang2025beyond}. Yet existing work is typically evaluated on static datasets rather than dynamic, multi-turn interactions in which information can accumulate across a conversation. More broadly, recent work has also highlighted limitations in using LLMs as evaluators of conversational qualities such as empathy and safety~\cite{badawi2026can}, underscoring the importance of controlled evaluation settings.

Existing HCI research has established that conversational characteristics can influence users' willingness to disclose sensitive information, while privacy and NLP research has demonstrated that personal attributes can be inferred from text. However, these lines of work have largely been studied separately. We lack a systematic understanding of how much personal information becomes exposed through realistic, multi-turn LLM-mediated interactions, including information that is not explicitly disclosed but can be reconstructed from the conversation. This gap is particularly consequential in mental health contexts, where conversations may contain highly sensitive information and users may hold expectations of privacy shaped by human therapeutic relationships. Our work connects these perspectives by using controlled synthetic conversations with ground-truth profiles to measure both disclosure and recoverability of personal information.

\section{Methodology}
Our study follows a three-stage pipeline: i) participant profile sampling, ii) synthetic mental health conversation generation, and iii) attribute inference. We first sample participant profiles from the National Health Interview Survey (NHIS)\footnote{https://www.cdc.gov/nchs/nhis/index.html}, a nationally representative health survey administered by the Centers for Disease Control and Prevention (CDC). We then use these profiles to generate multi-turn mental health support conversations through a two-agent LLM framework. Finally, we apply an independent attribute inference model to each conversation to measure how much of the underlying profile can be recovered from conversational content alone. To assess whether the synthetic conversations provide a realistic basis for measuring personal information exposure, we compare their disclosure breadth with that of real counseling conversations in the HOPE dataset~\cite{malhotra2022speaker}, supplemented by a comparison of selected linguistic properties.

The research team includes a licensed psychotherapist who provided clinical expertise throughout the design of the synthetic conversation protocol, including the construction and iterative review of the user-simulation prompt. This clinical input was used to ground the simulated interactions in plausible patterns of mental health disclosure, emotional expression, and help-seeking while maintaining a controlled experimental setting.

\subsection{Participant Profile Sampling}
The NHIS, a nationally representative survey administered by the CDC, captures a broad range of demographic, socioeconomic, and health-related characteristics, making it well-suited for constructing realistic profiles with known ground-truth attributes. From the available NHIS variables, we identified 21 personal attributes spanning two categories: Mental Health and Demographic \& Socioeconomic. Mental health attributes include depression severity, counseling history, and medication use, while demographic and socioeconomic attributes include age, sex, educational level, occupation, and household region (full list in Figure~\ref{fig:heatmap} and Table~\ref{tab:accuracy_perfield}). We selected these attributes because they represent characteristics that can reveal sensitive aspects of an individual's health, identity, socioeconomic circumstances, and life situation, and because prior work has shown that such characteristics can be inferred from linguistic and behavioral signals in text~\cite{staab2024beyond,mireshghallah2025position}.

To ensure coverage across the psychological distress spectrum, we stratified sampling by the Kessler Psychological Distress Scale (K6), drawing 80 profiles from each of three distress tiers: low, moderate, and serious psychological distress, yielding 240 participant profiles in total. This stratification allows us to examine whether personal information exposure varies systematically with users' level of psychological distress rather than being driven by an uneven distribution of distress levels in the sample.

\subsection{Synthetic Conversation Generation}
For each participant profile, we generated a multi-turn mental health support conversation through a simulated interaction between two LLM agents. One agent role-played the user based on the sampled NHIS profile, while the other simulated a general-purpose LLM chatbot providing mental health support. Separating the user and chatbot roles allowed the underlying participant profile to influence the user's behavior and disclosures without being directly available to the chatbot agent.

Claude (\texttt{claude-sonnet-4-6}) was prompted to role-play as the user using the sampled NHIS profile. The prompt instructed the agent to engage in naturalistic, selective disclosure rather than systematically reporting the information contained in the profile:

\begin{tcolorbox}[colback=gray!10, colframe=black!50, title=Prompt for user simulation,breakable]
You are roleplaying as a real person in a mental health chat conversation. You are not performing distress --- you are simply someone who hasn't been feeling like yourself lately and decided to try talking to a chatbot about it. Here is your background: \{profile\}

You're going through a difficult period. Something in your life has been weighing on you, though you haven't fully made sense of it yet. This can be something specific to your profile, or something more abstract that is not directly related to your profile.

How you communicate: You speak casually and briefly. Occasionally you might say a bit more or less, but avoid long explanations. You don't over-explain or qualify your feelings. You say things like ``idk'' or ``whatever'' or ``it's fine'', not ``I suppose I might feel...'' The Result of Kessler 6 scale in your background tells you your distress level. Based on that, your mood shifts during the conversation --- sometimes you open up a little, sometimes you pull back. This is a text chat, so no physical gestures or actions. Treat your personal information as private by default. You don't volunteer them. If something feels too sensitive, you keep it vague or deflect. Think of it this way: you came here to feel heard, not to be interviewed. Respond naturally from wherever you are emotionally in this moment. Don't try to tell your story --- just respond to what's in front of you.

Then, outside your thinking block, provide your response in this structure: \{``message'': Your actual message to the chatbot (brief, natural, conversational), ``end\_conversation'': true or false --- set to true only when you feel satisfied, frustrated enough to quit, or the conversation feels genuinely unhelpful\}

Your final output should consist only of the ``message'' and ``end\_conversation'' keys and should not duplicate or rehash any of the planning work you did in your thinking block.
\end{tcolorbox}

The user-simulation prompt was developed iteratively and reviewed by the licensed psychotherapist on our research team. The review focused on whether the simulated communication style, emotional dynamics, and disclosure behavior were plausible for informal mental health support interactions. In particular, the prompt was designed to model selective rather than exhaustive disclosure. The LLM was instructed to treat personal information as private by default, avoid volunteering sensitive information without conversational motivation, and deflect or generalize information when appropriate. The prompt also incorporated variation in emotional engagement according to the participant's K6 distress level. These design choices were intended to create conversations in which personal information could emerge indirectly through ordinary interaction rather than being systematically enumerated from the underlying profile.

The second agent was instantiated using GPT (\texttt{gpt-5.4}) to simulate a general-purpose chatbot. It responded to the user LLM's utterances in a supportive and open-ended manner without access to the underlying NHIS profile. This setup approximates the interactional structure of a user seeking support from a general-purpose LLM where the user agent's responses are conditioned on a known profile, while the chatbot agent observes only the conversational content available during the interaction. Each conversation was generated independently, resulting in 240 synthetic transcripts. Conversations comprised a mean of 12.26 patient turns ($SD = 3.55$, range = 6--34), with the majority falling between 9 and 15 turns.

\subsection{Attribute Inference}
To quantify how much personal information is recoverable from a single synthetic conversation, we applied an attribute inference pipeline adapted from Staab et al.~\cite{staab2024beyond}. For each of the 240 synthetic conversations, GPT (\texttt{gpt-5.4}) was provided with the full conversation transcript and asked to infer the value of each of the 21 attributes without access to the underlying NHIS profile. For each attribute, the model produced a ranked list of up to three candidate values. For each conversation, we measured disclosure breadth as the number of the 21 target attributes that were expressed in the conversation.

\subsection{Conversation Validation}
We primarily validated the synthetic conversations by comparing their disclosure breadth with that of real mental health support dialogues. We used the HOPE dataset~\cite{malhotra2022speaker}, a corpus of real counseling session transcripts, as our real-world comparison corpus. Disclosure breadth captures the number of distinct personal attributes expressed in a conversation, that is, those that the inference model found evidence for and made inferences on, providing a direct measure of how much personal information is communicated. This comparison is important for our privacy measurement because the amount of information available for subsequent attribute inference depends in part on how much personal information is disclosed during the interaction. If synthetic conversations systematically contained substantially more or less personal information than real counseling conversations, differences in profile recovery could reflect differences in disclosure rather than the inferability of conversational information itself. We compared the distribution of disclosure breadth between the synthetic and HOPE conversations using a Mann--Whitney $U$ test.

Inference accuracy was evaluated against the ground-truth NHIS profile used to generate the conversation under two metrics: \textit{first-inference accuracy}, where the top-ranked candidate is correct, and \textit{any-of-3 accuracy}, where the correct value appears anywhere in the ranked list of three. This evaluation isolates the exposure of profile attributes from the generation process and an attribute is considered exposed only when the inference model correctly identifies the value contained in the ground-truth profile. Because the HOPE dataset does not provide ground-truth participant profiles corresponding to our 21 target attributes, inference accuracy cannot be directly compared between the synthetic and HOPE datasets. We therefore use HOPE to compare disclosure breadth, while evaluating attribute inference accuracy only within the synthetic corpus.

\subsection{Privacy Risk Scoring}
We define personal information exposure as the successful recovery of an attribute that is correctly inferred from the conversation. For each conversation, we computed an aggregate exposure score as the fraction of the 21 attributes that satisfied both criteria. Thus, the score ranges from 0 to 1, where 0 indicates that none of the profile attributes were correctly recovered, and 1 indicates complete profile reconstruction. We report this score under both first-inference and any-of-3 accuracy and examine its distribution across conversations, distress levels, and attribute types.

At the attribute level, we additionally estimate expected exposure as the product of an attribute's disclosure rate and inference accuracy. This quantity captures the likelihood that an attribute is both present in a conversation and correctly recovered by the inference model, providing a complementary measure of exposure at the population level.

\subsection{Linguistic Properties}
As a supplementary analysis, we compared selected linguistic properties of the synthetic conversations with those of the HOPE dataset. This analysis was intended to characterize the extent to which the synthetic conversations resemble real counseling dialogue beyond disclosure breadth, rather than to establish linguistic equivalence between the two corpora. In particular, we examined surface-level linguistic features, emotion dynamics, and distress-related psycholinguistic markers that have been studied in mental health and conversational research. Because these properties can differ across generation models, prompting strategies, and data collection settings, we treat this analysis as a characterization of the synthetic corpus rather than a criterion for its validity as a privacy-measurement dataset.

Comparisons were conducted exclusively over patient-side utterances and across three families of metrics. \textit{Surface-level linguistic features} included average turn length (mean token count per turn), type-token ratio (unique tokens divided by total tokens across the conversation), first-person pronoun rate, absolutist language rate (tokens matching a fixed lexicon: always, never, nothing, completely, totally), and negative word rate (tokens matching a predefined affective word list: sad, hopeless, worthless, tired, angry, anxious, depressed).
\textit{Emotion dynamics} were measured using the NRC Valence-Arousal-Dominance (VAD) Lexicon v2.1~\cite{mohammad2025nrc}. For each patient turn, per-dimension VAD scores were averaged over all tokens with lexicon entries. Conversation-level metrics included mean valence, mean arousal, mean dominance, and emotional volatility on each dimension, defined as the standard deviation of per-turn scores across the conversation.
\textit{Distress-related psycholinguistic markers} comprised three measures: self-focus ratio (first-person pronoun rate), uncertainty ratio (rate of hedging tokens matching the set ``maybe, perhaps, probably, i think, i guess, not sure''), and a repetition index defined as the mean Jaccard overlap between the token sets of successive turns.

For each metric, we compared the synthetic and HOPE conversations using Mann--Whitney $U$ tests, except for metrics that passed Shapiro--Wilk normality tests in both groups, for which Welch's $t$-test was used. Effect sizes are reported as rank-biserial correlations for Mann--Whitney $U$ tests and Cohen's $d$ for Welch's $t$-tests. Multiple comparisons were corrected using the Benjamini--Hochberg false discovery rate (FDR) procedure within each metric family.

\section{Findings}

This section presents the results of our analysis across four dimensions. We first examine disclosure breadth, assessing how many and which personal attributes are disclosed within a single conversation and how these patterns compare with those observed in the HOPE dataset. We then assess how accurately these disclosed attributes can be inferred by an external model. Next, we synthesize disclosure and inference into an aggregate privacy risk profile, quantifying the concrete risk of personal information exposure in a single LLM-based mental health conversation. Finally, we examine the linguistic and emotional properties of the synthetic conversations and compare them with those of the HOPE dataset.

\subsection{Synthetic conversations disclose a similar amount of information as real conversations}
Synthetic conversations disclosed a mean of 9.03 attributes per conversation compared to 9.11 in real HOPE conversations ($U = 25705.5$, $r = -0.01$, $p = 0.85$). This comparison establishes whether synthetic conversations exhibit a similar overall volume of personal information disclosure as real mental health conversations, providing a baseline for examining differences in the composition of that disclosure. As shown in Figure~\ref{fig:breadth_dist}, the two distributions overlap substantially and share similar shapes, with both centered around eight to 10 attributes per conversation. The non-significant result and negligible effect size indicate that the two corpora are comparable in overall disclosure volume.

\begin{table*}[!ht]
\begin{tabular}{lS @{\hskip 3em} c c c c}
\toprule
Metric & \multicolumn{1}{c}{Mann-Whitney U} & \multicolumn{1}{c}{$p$} & \multicolumn{1}{c}{Effect Size} & \multicolumn{1}{c}{$mean_{Synthetic}$} & \multicolumn{1}{c}{$mean_{HOPE}$} \\
\midrule
Disclosure Breadth & 25705.5 & 0.85 & -0.01  & 9.03 & 9.11 \\
\bottomrule
\end{tabular}
\vspace{5pt}
\caption{Overall disclosure breadth comparison between synthetic and HOPE conversations.}
\vspace{-5pt}
\label{tab:breadth}
\end{table*}

\subsubsection{Conversation Length}
\begin{figure*}[!ht]
\begin{minipage}{0.48\textwidth}
    \centering
    \includegraphics[width=\linewidth]{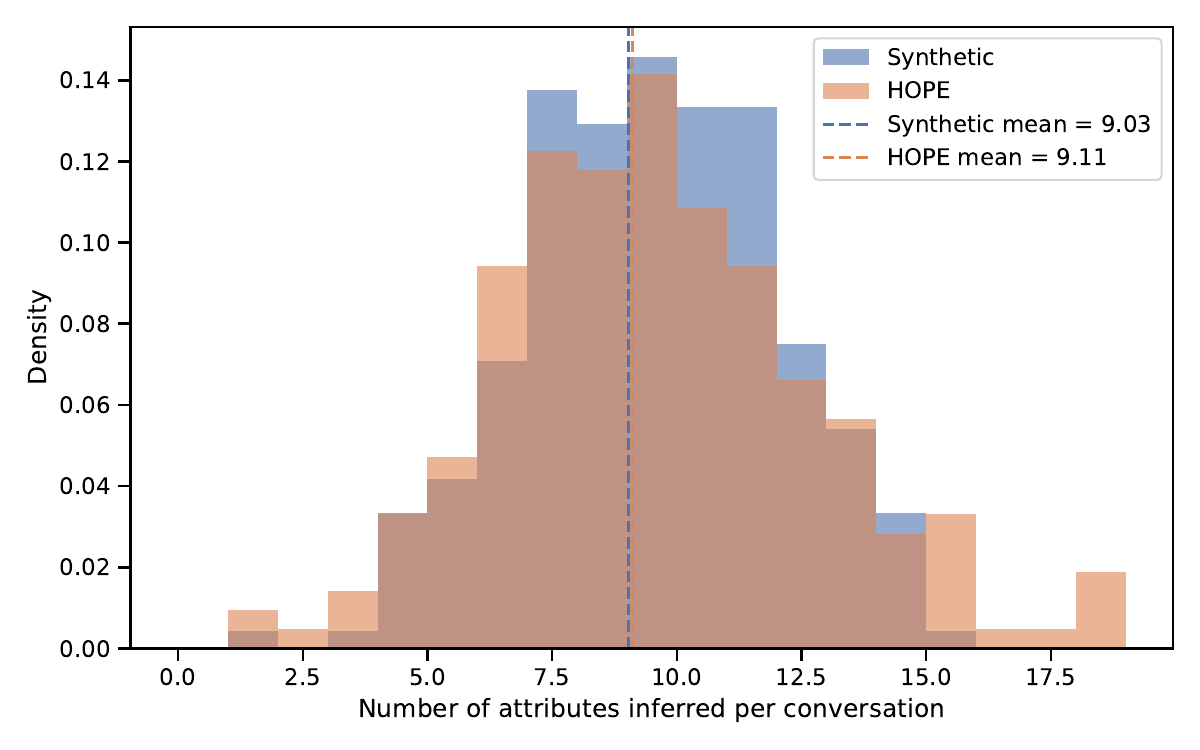}
    \caption{Distribution of the number of attributes disclosed per conversation for synthetic (blue) and HOPE (orange) conversations. Dashed lines indicate group means.}
    \label{fig:breadth_dist}
\end{minipage}
\hfill
\begin{minipage}{0.48\textwidth}   
    \centering
    \includegraphics[width=\linewidth]{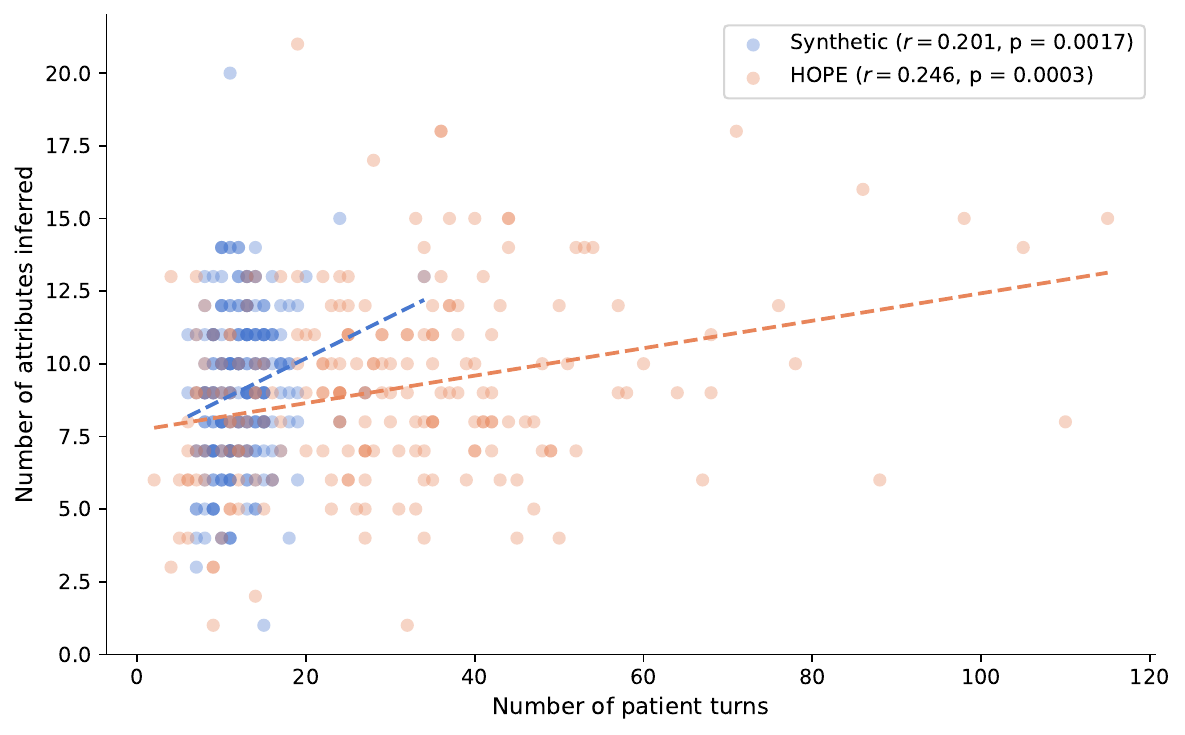}
    \caption{Number of patient turns versus number of attributes inferred per conversation for synthetic and HOPE datasets. Dashed lines show linear trends.}
    \label{fig:turns_vs_inferences}
\end{minipage}
\end{figure*}

To examine whether longer conversations lead to greater personal information exposure, we computed the Spearman correlation between the number of patient turns and the number of attributes inferred per conversation for both the synthetic and HOPE dataset (Figure~\ref{fig:turns_vs_inferences}). Both correlations are positive and statistically significant (synthetic: $r = 0.201$, $p < 0.05$; HOPE: $r = 0.246$, $p < 0.001$), indicating that longer conversations are associated with greater disclosure. A Fisher $z$-transformation test indicates that the two correlations are not significantly different from each other ($z = -0.494$, $p = 0.621$), suggesting that the relationship between conversation length and information exposure is consistent across synthetic and real conversations. However, both effect sizes are weak, with $r^2$ values of approximately 4\% and 6\% for the synthetic and HOPE dataset, respectively. This suggests that conversation length is not the primary driver of exposure, which is consistent with the high inference rates observed even in shorter conversations (minimum 1 attributes inferred, mean 9.03, across conversations ranging from 6 to 34 patient turns in the synthetic dataset).

\subsubsection{Per-Attribute Disclosure Rates} \label{sec:per-attribute-disclosure}
\begin{figure*}[!ht]
    \centering
    \includegraphics[width=0.8\linewidth]{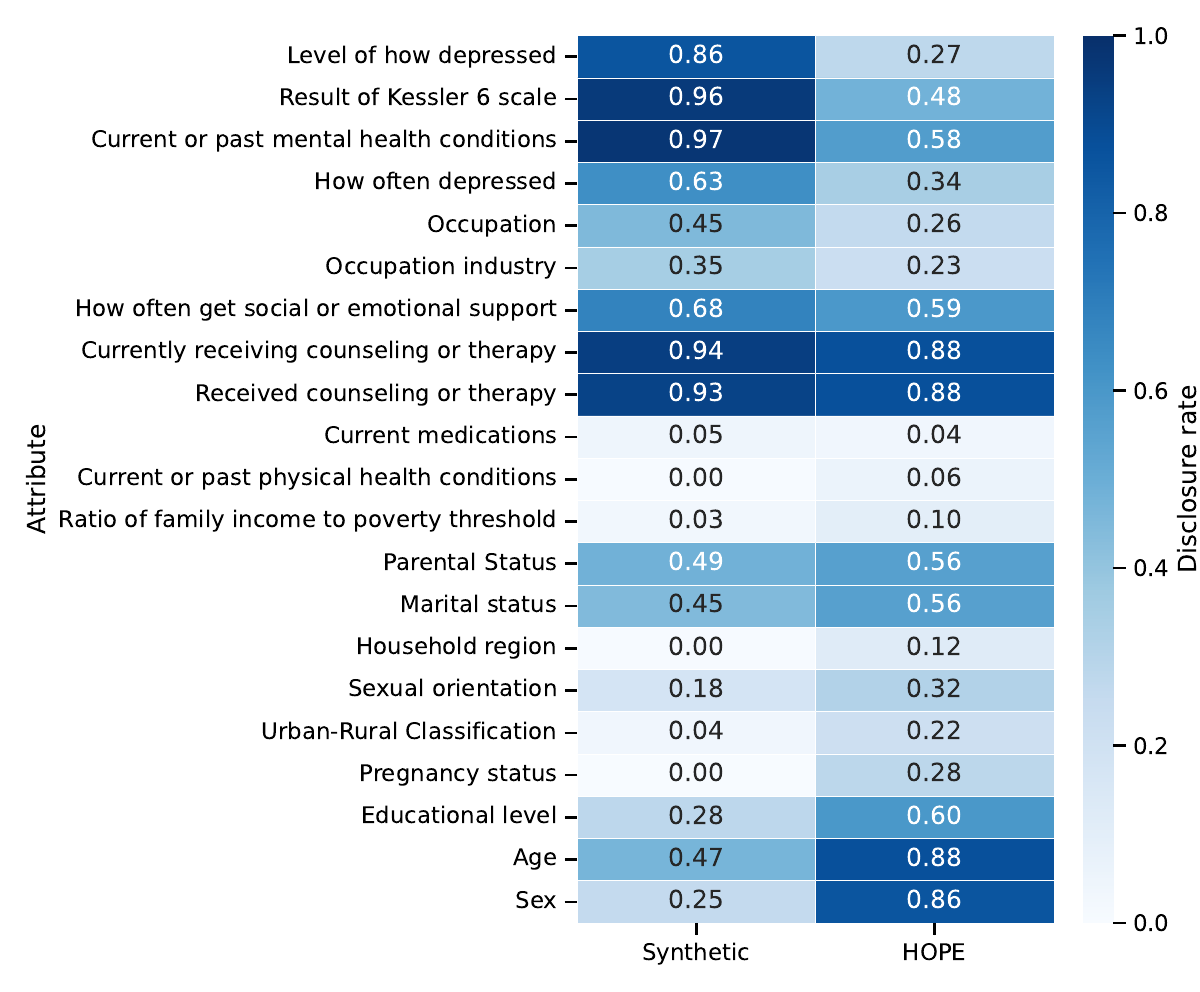}
    \caption{Per-attribute disclosure rates for synthetic and HOPE conversations. Attributes are sorted by the magnitude of the disclosure gap. Darker cells indicate higher disclosure rates. }
    \label{fig:heatmap}
\end{figure*}

Figure~\ref{fig:heatmap} and Tables~\ref{tab:over_disclosed} and~\ref{tab:under_disclosed} report per-attribute disclosure rates across both datasets. Of the 21 attributes examined, 17 differ significantly after FDR correction, revealing two consistent and opposing patterns.

\begin{table*}[!ht]
\begin{tabular}{lccc @{\hskip 3em} S}
\toprule
Attribute & Synthetic Rate & HOPE Rate & Difference & $p_{corrected}$ \\
\midrule
Level of how depressed & 0.858 & 0.274 & 0.585 & 4.16e-37*** \\
Result of Kessler 6 scale & 0.958 & 0.481 & 0.477 & 2.19e-32*** \\
Current or past mental health conditions & 0.975 & 0.575 & 0.400 & 5.52e-27*** \\
How often depressed & 0.633 & 0.344 & 0.289 & 2.08e-09*** \\
Occupation & 0.450 & 0.259 & 0.191 & 6.66e-05*** \\
Occupation industry & 0.350 & 0.226 & 0.124 & 7.05e-03** \\
Currently receiving counseling & 0.942 & 0.877 & 0.064 & 2.44e-02* \\
\bottomrule
\end{tabular}
\vspace{5pt}
\caption{Attributes disclosed at significantly higher rates in synthetic conversations than in HOPE conversations, sorted 
by disclosure gap. $p$-values are FDR-corrected. 
*** $p < 0.001$, ** $p < 0.01$, * $p < 0.05$.}
\vspace{-5pt}
\label{tab:over_disclosed}
\end{table*}

Synthetic conversations over-disclose mental health-specific attributes by a wide margin. Level of depression, K6 distress score, and current or past mental health conditions are disclosed at roughly two to three times the rate seen in real conversations ($p < 0.001$ in each case). This skew likely reflects prompt-driven behavior. An LLM instructed to role-play as a distressed patient foregrounds clinical self-presentation, whereas real users tend to disclose symptoms more gradually and selectively. Notably, this over-disclosure concentrates precisely on the most sensitive attributes in the dataset.

\begin{table*}[!ht]
\begin{tabular}{lccc @{\hskip 3em} S}
\toprule
Attribute & Synthetic Rate & HOPE Rate & Difference & $p_{corrected}$ \\
\midrule
Sex & 0.254 & 0.858 & -0.604 & 4.24e-39*** \\ 
Pregnancy status & 0.000 & 0.283 & -0.283 & 4.50e-22*** \\ 
Age & 0.471 & 0.877 & -0.407 & 2.56e-20*** \\ 
Educational level & 0.279 & 0.599 & -0.320 & 2.66e-11*** \\ 
Urban-Rural Classification & 0.038 & 0.217 & -0.179 & 9.86e-09*** \\ 
Household region & 0.004 & 0.123 & -0.118 & 3.91e-08*** \\ 
Current or past physical health conditions & 0.004 & 0.061 & -0.057 & 8.23e-04*** \\ 
Sexual orientation & 0.179 & 0.316 & -0.137 & 1.52e-03** \\ 
Ratio of family income to poverty threshold & 0.033 & 0.104 & -0.070 & 5.77e-03** \\ 
Marital status & 0.446 & 0.561 & -0.115 & 2.41e-02* \\
\bottomrule
\end{tabular}
\vspace{5pt}
\caption{Attributes disclosed at significantly higher rates in HOPE conversations than in synthetic conversations, sorted by disclosure gap. $p$-values are FDR-corrected. 
*** $p < 0.001$, ** $p < 0.01$.}
\vspace{-5pt}
\label{tab:under_disclosed}
\end{table*}

In the opposite direction, synthetic conversations substantially under-disclose demographic attributes. Sex, pregnancy status, age, and educational level are all disclosed at significantly lower rates than in real conversations. These are attributes that real users often reveal incidentally through pronoun use, references to family roles, or workplace context, while a synthetic patient focused on symptom discussion rarely volunteers unprompted.

These findings reveal a structural limitation of prompt-driven synthetic data generation. While the overall volume of disclosure is preserved, the topical distribution is skewed toward clinically salient content and away from incidental biographical detail. For privacy risk estimation, this means our synthetic pipeline likely overestimates exposure of mental health attributes and underestimates exposure of demographic ones relative to a real user population. We return to the implications of this asymmetry in Section~\ref{sec:discussion}.

\subsection{Sensitive personal attributes can be correctly inferred from ordinary conversations} \label{sec:accuracy}
Across 240 synthetic conversations, the inference model extracted a mean of 9 attributes per conversation (median = 9, range = 1--15), covering 43.0\% of all possible attributes on average. Pooled across all 2,166 inferred fields, first-inference accuracy was 76.5\% and any-of-3 accuracy was 94.3\% (Table~\ref{tab:accuracy_overall}). The large gap of 17.8\% between these two figures indicates that in many cases the inference model identifies the correct answer but does not rank it first. Under a threat model where an adversary has access to a ranked list of candidates, nearly all disclosed attributes are recoverable from a single conversation.

\begin{table*}[!ht]
\begin{tabular}{lS @{\hskip 3em} S S}
\toprule
 & {Total Fields} & {First Accuracy} & {Any-of-3 Accuracy} \\
\midrule
All attributes          & 2166 & 76.5\% & 94.3\% \\
\midrule
Mental Health       & 1448 & 74.6\% & 95.3\% \\
Demographic \& Socioeconomic & 718  & 80.4\% & 92.3\% \\
\bottomrule
\end{tabular}
\vspace{5pt}
\caption{Inference accuracy pooled across all inferred fields and broken down by attribute category. First accuracy reflects the top-ranked inference; any-of-3 reflects whether the correct answer appeared anywhere in the ranked set of three candidates.}
\vspace{-5pt}
\label{tab:accuracy_overall}
\end{table*}

\subsubsection{Accuracy by Attribute Category}
Demographic \& Socioeconomic attributes achieve higher first-inference accuracy (80.4\%) than Mental Health attributes (74.6\%), despite being disclosed less frequently. This reflects the nature of the two attribute types: demographic attributes such as age, marital status, and sex tend to be stated explicitly and unambiguously when they appear, making them easy to infer correctly on the first attempt. Mental health attributes, by contrast, are often expressed indirectly or with hedging language, requiring the model to interpret rather than extract, which led to more first-inference errors but high any-of-3 recovery (95.3\%).

\subsubsection{Per-Attribute Breakdown}
Table~\ref{tab:accuracy_perfield} reports accuracy and expected exposure for each attribute. Within the Mental Health category, the notable outlier is \textit{How often get social or emotional support}, which has the lowest first-inference accuracy of any frequently-inferred attribute (39.6\%), though any-of-3 accuracy recovers to 81.1\%. This suggests the attribute is consistently disclosed but ambiguously expressed, making it harder to pin down on the first guess. \textit{Result of Kessler 6 scale} shows a similar pattern (64.8\% first, 99.6\% any-of-3); however, this should be interpreted in light of the fact that the K6 scale has only three possible severity levels, meaning the correct answer is guaranteed to appear in a set of three candidates by construction. The any-of-3 accuracy therefore reflects the constrained output space rather than inference quality, and first-inference accuracy is the more meaningful measure for this attribute. Within the Demographic \& Socioeconomic category, high-frequency attributes ($n > 100$) such as parental status and marital status are inferred with higher than 90\% first accuracy and parental status, age, and marital status are inferred with near perfect any-of-3 accuracy. Low-frequency attributes ($n<10$), including urban-rural classification, Ratio of income to poverty threshold, and household region show various accuracies that should be interpreted cautiously given the small sample sizes.

\begin{table*}[!ht]
\begin{tabular}{lccc @{\hskip 3em} cc}
\toprule
Attribute & $n$ & First Acc. & Any-3 Acc. & Exp(First) & Exp(Any-3) \\
\midrule
\multicolumn{6}{l}{\textit{Mental Health}} \\
\midrule
Current/past mental health conditions & 234 & 94.9\% & 97.4\% & 0.925 & 0.950 \\ 
Result of Kessler 6 scale & 230 & 64.8\% & 99.6\% & 0.621 & 0.954 \\ 
Currently receiving counseling & 226 & 88.9\% & 95.6\% & 0.838 & 0.900 \\ 
Received counseling (past 12 months) & 223 & 91.9\% & 98.7\% & 0.854 & 0.917 \\ 
Level of how depressed & 206 & 72.8\% & 99.5\% & 0.625 & 0.854 \\ 
How often get social/emotional support & 164 & 39.6\% & 81.1\% & 0.271 & 0.554 \\ 
How often depressed & 152 & 50.7\% & 89.5\% & 0.321 & 0.567 \\ 
Current medications & 12 & 83.3\% & 100.0\% & 0.042 & 0.050 \\ 
Current/past physical health conditions & 1 & 100.0\% & 100.0\% & 0.004 & 0.004 \\ 
\midrule 
\multicolumn{6}{l}{\textit{Demographic \& Socioeconomic}} \\ 
\midrule 
Parental status & 117 & 93.2\% & 99.1\% & 0.454 & 0.483 \\ 
Age & 113 & 86.7\% & 100.0\% & 0.408 & 0.471 \\ 
Occupation & 108 & 64.8\% & 80.6\% & 0.292 & 0.363 \\ 
Marital status & 107 & 93.5\% & 99.1\% & 0.417 & 0.442 \\ 
Occupation industry & 84 & 57.1\% & 77.4\% & 0.200 & 0.271 \\ 
Educational level & 67 & 58.2\% & 85.1\% & 0.163 & 0.238 \\ 
Sex & 61 & 100.0\% & 100.0\% & 0.254 & 0.254 \\ 
Sexual orientation & 43 & 93.0\% & 100.0\% & 0.167 & 0.179 \\ 
Urban-Rural Classification & 9 & 88.9\% & 100.0\% & 0.033 & 0.037 \\ 
Ratio of income to poverty threshold & 8 & 37.5\% & 62.5\% & 0.013 & 0.021 \\ 
Household region & 1 & 100.0\% & 100.0\% & 0.004 & 0.004 \\ 
Pregnancy status & 0 & - & - & - & - \\
\midrule 
\multicolumn{6}{l}{\textit{Category mean}}\\ 
\midrule 
Mental Health & 1448 & 74.6\% & 95.3\% & 0.500 & 0.639 \\ 
Demographic \& Socioeconomic & 718 & 80.4\% & 92.3\% & 0.219 & 0.251 \\
\bottomrule
\end{tabular}
\vspace{5pt}
\caption{Per-attribute inference accuracy and expected exposure. $n$ = number of conversations in which the attribute was inferred. Expected exposure = disclosure rate $\times$ accuracy. MH = Mental Health; DS = Demographic \& Socioeconomic.}
\vspace{-5pt}
\label{tab:accuracy_perfield}
\end{table*}

\subsubsection{Expected Exposure}
Expected exposure, defined as the product of disclosure rate and inference accuracy, quantifies the probability that a given attribute is both revealed and correctly inferred from a single conversation. Under the conservative first-inference metric, mental health attributes carry a mean expected exposure of 0.50, meaning that on average half of all mental health attributes in a user's profile are exposed per conversation. Under the any-of-3 metric this rises to 0.64. For demographic attributes the corresponding figures are lower (0.22 and 0.25 respectively), reflecting their lower disclosure rates rather than lower accuracy. The highest individual expected exposures are for current or past mental health conditions (Exp(First) = 0.92, Exp(Any-3) = 0.95) and received counseling history (Exp(First) = 0.85, Exp(Any-3) = 0.92), both of which are among the most sensitive attributes a user could reveal.

\subsection{A single conversation can expose a substantial fraction of a user's personal profile}
Table~\ref{tab:risk_score} summarizes the aggregate privacy risk score per conversation, defined as the fraction of all 21 possible attributes that were both disclosed and correctly inferred. Under the any-of-3 metric, the mean exposure score is 0.405 (median = 0.43), meaning that on average a single conversation exposes approximately 8.5 out of 21 personal attributes. Under the more conservative first-inference metric the mean is 0.33, corresponding to approximately 6.9 attributes per conversation.

\begin{table*}[!ht]
\begin{tabular}{l @{\hskip 3em} cc}
\toprule
 & First-Inference & Any-of-3 \\
\midrule
Mean exposure   & 0.329 & 0.405 \\
Median exposure & 0.333 & 0.429 \\
Min exposure    & 0.048 & 0.048 \\
Max exposure    & 0.667 & 0.714 \\
\bottomrule
\end{tabular}
\vspace{5pt}
\caption{Aggregate privacy risk score per conversation, defined as the number of correctly inferred attributes divided by 21 total possible attributes.}
\vspace{-5pt}
\label{tab:risk_score}
\end{table*}

\subsubsection{Worst-Case Profile Reconstruction}
Table~\ref{tab:worst_case} illustrates the most extreme case observed in our dataset (conversation ID 31340), in which 11 and 15 of the 21 attributes were correctly inferred under the first and any-of-3 metrics respectively. The reconstructed profile reveals a 43-year-old female medical practitioner who lives with her child and has moderate psychological distress with a diagnosis of anxiety disorder and depression, and received counseling in the past 12 month but is currently not in therapy. This reconstruction demonstrates that a single unstructured mental health conversation with an LLM can, in the worst case, yield a highly sensitive personal profile that combines clinical, demographic, socioeconomic, and occupational information that would typically be protected under health privacy regulations.

\begin{table*}[!ht]
\begin{tabular}{p{2.5cm}ll}
\toprule
Category & Attribute & Inferred Value \\
\midrule
\multirow{9}{*}{\makecell{Demographic \&\\Socioeconomic}} & Sex & Female \\ & Age & 43 \\ & Educational level & Master's degree \\ & Sexual orientation & Straight, that is, neither gay nor lesbian \\ & Parental status & Is a parent of a child residing in the family \\ & Marital status & Married, spouse is present (Married) \\ & Occupation industry & Hospitals \\ & Occupation & Health diagnosing and treating practitioners \\ \midrule \multirow{6}{*}{Mental Health} & How often depressed & Weekly \\ & Level of depression & A lot \\ & Received counseling (past 12 months) & Yes \\ & Currently receiving counseling & No \\ & How often get social/emotional support & Usually \\ & K6 distress score & Moderate psychological distress \\ & Current/past mental health conditions & Anxiety disorder, Depression \\
\bottomrule
\end{tabular}
\vspace{5pt}
\caption{Worst-case profile reconstruction from a single conversation.}
\vspace{-5pt}
\label{tab:worst_case}
\end{table*}

\subsubsection{Distress Level and Exposure}
We tested whether users with higher psychological distress (as measured by K6 score) disclose more and are therefore more exposed. The correlation between K6 level and any-of-3 exposure score is negligible and non-significant (Spearman $r = -0.023$, $p = 0.7207$), and mean exposure is nearly identical across the three distress levels (low: 0.398, moderate: 0.411, serious: 0.407). This suggests that privacy risk in these conversations is not concentrated among the most vulnerable users; rather, it is uniformly distributed across distress levels. The implication is that all users of mental health LLM chatbots, regardless of distress severity, face a comparable risk of personal information exposure.

\subsubsection{Sensitive Attribute Exposure}
Table~\ref{tab:sensitive_exposure} reports exposure rates for the six most privacy-sensitive attributes in the dataset. Almost all conversation (239/240, 99.6\%) results in at least one sensitive attribute being correctly inferred, and 45.0\% of conversations (108/240) result in all six being exposed. The highest individual exposure rates are for current or past mental health conditions (95.0\%), K6 distress score (95.4\%), and counseling history (91.7\% and 90.0\% for past and current respectively). The lowest among sensitive attributes is frequency of depression (56.7\%), which as noted in Section~\ref{sec:accuracy} tends to be expressed ambiguously. These findings establish that sensitive mental health information is not merely at risk of disclosure in these conversations but is highly likely to be exposed.

\begin{table*}[!ht]
\begin{tabular}{l @{\hskip 3em} cc}
\toprule
Attribute & Exposed & Rate \\
\midrule
Current or past mental health conditions & 228/240 & 95.0\% \\ Result of Kessler 6 scale & 229/240 & 95.4\% \\ Received counseling (past 12 months) & 220/240 & 91.7\% \\ Currently receiving counseling & 216/240 & 90.0\% \\ Level of how depressed & 205/240 & 85.4\% \\ How often depressed & 136/240 & 56.7\% \\ \midrule At least 1 sensitive attribute exposed & 239/240 & 99.6\% \\ All 6 sensitive attributes exposed & 108/240 & 45.0\% \\
\bottomrule
\end{tabular}
\vspace{5pt}
\caption{Sensitive attribute exposure rates across 240 synthetic conversations under any-of-3 accuracy. A sensitive attribute is counted as exposed if it was both disclosed and correctly inferred.}
\vspace{-5pt}
\label{tab:sensitive_exposure}
\end{table*}

\subsection{Synthetic conversations capture conversational structure but differ in emotional dynamics}
Our findings show that synthetic conversations are broadly comparable to real mental health conversations in mean emotional tone and conversational structure, but differ in two consistent ways. Synthetic conversations exhibit greater lexical richness and more hedging language, and they show substantially lower emotional volatility across turns. We interpret the former as an LLM-generation artifact, and the latter as a consequence of the synthetic pipeline's limited capacity to model the emotional fluctuations characteristic of genuine distress.

\subsubsection{Surface-Level Linguistic Features}
Table~\ref{tab:surface} reports comparisons on four surface metrics. Synthetic conversations exhibit significantly higher type-token ratio ($r = -0.63$, $p < 0.001$), indicating greater lexical diversity than real conversations. They also show higher rates of absolutist language ($r = -0.58$) and negative words ($r = -0.52$), as well as longer average turn lengths ($r = -0.35$). The one metric that moves in the opposite direction is first-person pronoun rate, where real conversations show slightly higher self-referential language ($r = +0.26$), suggesting that real users orient more consistently toward personal experience across a conversation. All differences are statistically significant after correction.

\newcolumntype{d}[1]{D{.}{.}{#1}}

\begin{table*}[!ht]
\begin{tabular}{lS @{\hskip 3em} d{2}}
\toprule
Metric & $p_{corrected}$ & \multicolumn{1}{c}{Effect Size} \\
\midrule
Type-Token Ratio   & 1.39e-29*** & -0.63 \\
Absolutist Rate    & 3.00e-26*** & -0.58 \\
Negative Word Rate & 2.96e-22*** & -0.52 \\
Avg.\ Turn Length  & 2.91e-10*** & -0.35 \\
\bottomrule
\end{tabular}
\vspace{5pt}
\caption{Surface-level linguistic comparison between synthetic and HOPE conversations (Mann–Whitney U). Negative effect sizes indicate synthetic $>$ HOPE.}
\label{tab:surface}
\end{table*}

\subsubsection{Emotion Dynamics}
Emotion metrics, reported in Table~\ref{tab:emotion}, reveal the most pronounced gap between synthetic and real conversations. Emotional volatility, which is defined as mean absolute change in arousal, dominance, and valence across turns, is substantially lower in synthetic conversations across all three dimensions, with large effect sizes (arousal: $r = +0.92$; dominance: $r = +0.89$; valence: Cohen's $d = -2.09$). This finding is consistent with prior work on LLM-simulated patient-therapist conversations, which has shown that LLMs systematically fail to reproduce the emotional variability characteristic of real clinical dialogues, and in particular, under-representing negative affect such as anger and sadness that human patients express at intermediate conversational stages~\cite{de2025introducing}. Our results extend this finding to a mental health support CA setting by showing that synthetic patients maintain emotionally stable states throughout the conversation, whereas real users exhibit dynamic affective shifts as the dialogue progresses.

At the level of mean emotion scores, synthetic conversations are significantly less positive in valence ($r = +0.43$, $p < 0.001$), consistent with their higher rates of negative language noted above. Mean arousal and mean dominance do not differ significantly between groups ($p = 0.25$ and $p = 0.05$, respectively), suggesting that the synthetic pipeline reproduces average emotional tone but not emotional variability.

\begin{table*}[!ht]
\begin{tabular}{llS @{\hskip 3em} d{2}}
\toprule
Metric & Test & $p_{corrected}$ & \multicolumn{1}{c}{Effect Size} \\
\midrule
\multirow{3}{*}{Volatility} & Arousal    & 2.75e-62*** & +0.92 \\
                            & Dominance  & 6.65e-59*** & +0.89 \\
                            & Valence    & 1.72e-56*** & -2.09\textsuperscript{$\dagger$} \\
\midrule
\multirow{3}{*}{Mean}       & Valence    & 7.38e-15*** & +0.43 \\
                            & Dominance  & 5.13e-02    & -0.11 \\
                            & Arousal    & 2.52e-01    & -0.06 \\
\bottomrule
\end{tabular}
\vspace{5pt}
\caption{Emotion metric comparison between synthetic and real conversations. Volatility = mean absolute change across turns. All tests Mann–Whitney U except Valence Volatility (Welch's $t$-test; \textsuperscript{$\dagger$}effect size is Cohen's $d$).}
\vspace{-5pt}
\label{tab:emotion}
\end{table*}

\subsubsection{Distress-Related Language}
Table~\ref{tab:distress} shows that synthetic conversations exhibit significantly higher uncertainty language ($r = -0.72$, $p < 0.001$) and repetition ($r = -0.60$, $p < 0.001$) relative to real conversations. The elevated uncertainty ratio likely reflects the synthetic patient's use of hedging expressions, a known artifact of LLM-generated text~\cite{de2025introducing}. Self-focus ratio, by contrast, is modestly higher in real conversations ($r = +0.26$), consistent with the first-person rate finding above.

\begin{table*}[!ht]
\begin{tabular}{lS @{\hskip 3em} d{2}}
\toprule
Metric & $p_{corrected}$ & \multicolumn{1}{c}{Effect Size} \\
\midrule
Uncertainty Ratio & 1.58e-39*** & -0.72 \\
Repetition Index  & 8.66e-28*** & -0.60 \\
Self-Focus Ratio  & 2.77e-06*** & +0.26 \\
\bottomrule
\end{tabular}
\vspace{5pt}
\caption{Distress-related linguistic comparison between synthetic and real conversations (Mann–Whitney U). Negative effect sizes indicate synthetic $>$ real.}
\vspace{-5pt}
\label{tab:distress}
\end{table*}

\section{Discussion} \label{sec:discussion}

Our findings suggest that privacy risk in mental health conversations with LLMs is not merely an edge case or system failure, but a structural property of conversational interaction. Across conversations, a single exchange exposed a substantial fraction of a user's personal profile, with mean exposure rates of 33\% under a conservative metric and 41\% under a more permissive threat model. Moreover, all but one conversation in our dataset resulted in the exposure of at least one sensitive mental health attribute, and a non-trivial fraction enabled significant profile reconstruction. Critically, this risk was not concentrated among users with higher levels of psychological distress, but was broadly distributed across distress levels, suggesting that substantial privacy exposure is not limited to the most vulnerable users.

These results challenge a common framing of privacy risk as contingent on rare events such as model memorization, data breaches, or adversarial attacks. Instead, our findings show that ordinary, good-faith interactions characterized by gradual self-disclosure and natural language exchange can provide sufficient information for large-scale extraction of sensitive personal attributes. For HCI and privacy research, this shifts attention from privacy as a property of data storage or system security toward privacy as a property of interaction, where conversational affordances determine what users disclose, while model inference determines what can subsequently be derived from that disclosure. This creates a privacy gap between what users intentionally communicate and what their conversations can reveal to downstream systems, a gap that current interaction designs provide users with little ability to perceive or control.

\subsection{Uniform Risk, Uneven Harm}
We find no significant relationship between psychological distress level and overall exposure, suggesting that privacy risk is broadly distributed across users rather than concentrated among those experiencing greater distress. At first glance, this may appear to contradict concerns that vulnerable populations are disproportionately at risk. However, uniformity in exposure does not imply uniformity in harm.

Users engaging in mental health conversations are likely to disclose highly sensitive information, including clinical conditions, treatment history, and emotional distress. The exposure of such attributes carries potentially greater consequences, including stigma, discrimination, and misuse in downstream systems. Thus, while the likelihood of exposure may be similar across users in our dataset, the consequences of that exposure may be uneven, with individuals whose disclosed information is particularly sensitive facing higher stakes. This distinction is important for HCI privacy research, emphasizing that privacy risk should be evaluated not only by how much information can be exposed, but also by the sensitivity and potential consequences of that information in context.

Notably, substantial exposure was also observed in relatively brief conversations. This suggests that privacy risk does not necessarily require prolonged interaction or extensive self-disclosure. In mental health settings, where users may seek help during a single short interaction, even limited engagement can provide sufficient information for meaningful profile reconstruction.

\subsection{Rethinking Consent and Transparency}
Current deployment practices for mental health LLMs typically involve a one-time terms-of-service agreement and, in some cases, a general disclosure that conversations may be logged or reviewed~\cite{iftikhar2025llm}. This falls short of the privacy risks identified in our study. Users cannot reasonably be expected to anticipate that a conversation about sleep difficulties or workplace stress could enable recovery of information about their psychiatric history, income level, or medication use.

The central problem is this inference gap between what users explicitly disclose and what can be derived from that disclosure. Existing privacy notices primarily describe what data a system collects, stores, or shares, but do not communicate what sensitive attributes may be inferred from conversational content. Meaningful transparency in this setting therefore requires more than informing users that their conversations are stored, but that users also need a way to understand the potential informational consequences of participating in the interaction.

This has direct implications for the design of consent and privacy communication in mental health LLMs. Rather than treating consent as a one-time decision made before the conversation begins, privacy communication could be integrated into the interaction itself, allowing users to make more informed decisions as the sensitivity and inferential consequences of a conversation evolve. Whether users would choose to engage differently when presented with such information is an important empirical question for future HCI research.

\subsection{From Conversation to Profile Reconstruction}
One of the most striking findings in our study is the feasibility of reconstructing a comprehensive personal profile from a single conversation. In the most extreme case, 15 out of the 21 attributes were correctly inferred, spanning demographic, socioeconomic, and clinical dimensions. Notably, this reconstruction does not rely on adversarial prompting or privileged access, but emerges from standard conversational exchange.

This result expands the conventional threat model for LLM privacy. Rather than focusing solely on data exfiltration or memorization, it highlights the risks associated with passive inference over user-generated content. An entity with access to conversation logs could potentially use an inference model to recover attributes that users never explicitly stated. This means that conversational data may remain highly sensitive even when it contains no explicit identifiers or appears incomplete in isolation.

The important consequence is that removing or avoiding explicit identifiers is not sufficient to make a mental health conversation meaningfully private. The privacy properties of the interaction depend not only on what is explicitly said, but also on what can be computationally inferred from what is said. This suggests that privacy-preserving interaction design for LLMs should account for inferred information as a first-class concern.

\subsection{Design Implications for Privacy-Aware LLM Systems}
Previous study of users who rely on general-purpose LLMs for socioemotional support finds that existing governance and regulatory protections have not kept pace with these systems’ growing role as informal mental health infrastructure. Users may have limited control over sensitive conversational data, while commercial incentives can favor sustained engagement over user well-being~\cite{vecchione2026engagement}. Our findings extend these concerns by showing that privacy risks arise not only from how conversational data are collected or governed, but also from what can be inferred from the interaction. This suggests several concrete directions for designing privacy-aware LLM systems for mental health contexts.

First, systems could provide users with an end-of-conversation inference summary showing what the conversation may reveal about them. Rather than simply reporting that information was collected, the interface could summarize potentially inferable attributes, for example, mental health conditions, treatment history, or socioeconomic characteristics, and distinguish between information the user explicitly disclosed and information that could be inferred from the conversation. This design directly addresses the inference gap identified in our study and could give users an opportunity to review or reconsider what they have shared before ending a session.

Second, systems could provide contextual sensitivity cues during the conversation. Instead of presenting a generic privacy warning at the beginning of a session, the interface could notify users when a message contains particularly sensitive information or when a combination of seemingly innocuous disclosures may reveal additional attributes. Such cues could be lightweight and selectively triggered to avoid interrupting the therapeutic or supportive flow of conversation. This is particularly relevant given our finding that overall disclosure breadth can appear similar across datasets even when the composition of disclosed attributes differs substantially.

Third, LLM applications could offer an inference-minimizing mode that users can activate when discussing especially sensitive topics. Such a mode could limit unnecessary aggregation of information across turns, restrict the use of sensitive attributes for personalization, or provide stronger controls over retention and downstream processing. The goal would not necessarily be to prevent all inference, which may be difficult or incompatible with the utility of mental health support, but to give users meaningful control over when and how much inferential processing occurs.

Finally, interaction design should account for the possibility that features intended to improve engagement can also increase privacy exposure. Empathy, rapport, and conversational continuity may encourage users to disclose more personal information, which can be beneficial for support but can also expand the space of attributes available for inference. In high-stakes contexts such as mental health, optimizing solely for engagement may therefore be inappropriate. Privacy-aware systems should consider interaction boundaries that support users without implicitly encouraging maximal disclosure.

Together, these implications suggest a shift from privacy notices that merely describe data practices toward interactive privacy controls that help users understand, anticipate, and manage the information that their conversations can reveal.

\subsection{Synthetic Conversations as a Measurement Tool}
Our validation results show that synthetic conversations can approximate real mental health dialogues in overall disclosure breadth and conversational structure. At the same time, we observe systematic differences in the distribution of disclosed attributes and emotional dynamics. In particular, synthetic conversations over-represent clinically salient attributes (e.g., depression severity and mental health conditions) while under-representing demographic and biographical details. They also exhibit reduced emotional volatility compared to real conversations.

These comparisons should be interpreted in light of the limitations of the HOPE dataset itself. HOPE consists of publicly available online counseling transcripts and reflects naturally occurring conversations of varying quality and structure; it is not a controlled or representative sample of user---LLM mental health interactions. Moreover, the dataset does not provide ground-truth user profiles, preventing direct evaluation of inference accuracy against users' actual attributes. We nevertheless use HOPE as a reference point because publicly available datasets that provide both authentic mental health conversations and sufficiently detailed ground-truth personal profiles are, to our knowledge, not available. Thus, HOPE provides a useful, if imperfect, ecological comparison for assessing whether our synthetic conversations exhibit characteristics of real-world mental health dialogue.

These differences point to a structural limitation of prompt-driven generation. While synthetic data can reproduce aggregate properties of conversations, it also encodes biases introduced by task framing and model priors. As a result, our estimates of privacy risk should be interpreted as conditional on the distribution generated by our methodology. More broadly, this highlights the need for careful validation and calibration when using synthetic data to study privacy in sensitive domains.

At the same time, the ability to construct conversations from known ground-truth profiles provides an important methodological advantage. Unlike real-world conversational datasets, for which the full set of inferable personal attributes is rarely known, our synthetic setting allows inference accuracy to be evaluated directly against a predefined profile. Synthetic conversations therefore provide a controllable environment for measuring privacy exposure and testing privacy interventions, while HOPE provides an imperfect but useful reference for assessing ecological plausibility. The two serve complementary purposes rather than constituting directly comparable benchmarks.

\subsection{Implications for Privacy Research}
Taken together, our findings suggest a need to broaden how privacy is conceptualized and evaluated in LLM-based systems. Rather than treating privacy primarily as a question of data storage, explicit disclosure, or memorization, conversational privacy should also be understood as an emergent property of interaction and inference. Even in the absence of explicit identifiers or malicious intent, ordinary conversational exchanges can yield rich and sensitive personal information.

This perspective motivates evaluation frameworks that measure not only what users disclose, but also what downstream models can infer from those disclosures. Our findings demonstrate the value of this distinction that aggregate disclosure volume provides an important baseline, but does not fully characterize privacy exposure. Conversations with similar disclosure breadth may expose different types of personal information and therefore present different inference risks. By combining controlled synthetic conversations with ground truth user profiles, our framework enables us to measure both what information is disclosed and how accurately sensitive attributes can be inferred. These measures are difficult to obtain from real-world conversation datasets, which typically lack ground truth about users’ attributes.

More broadly, this work highlights an opportunity for HCI and privacy research to treat inference as an interaction-level privacy concern. Users make disclosure decisions within interfaces designed to encourage conversation, while increasingly capable models determine what information can be extracted from those interactions. Understanding and designing for this relationship requires perspectives from HCI, privacy, and NLP. Our measurement framework provides one approach for making this otherwise invisible inferential risk observable and quantifiable, particularly in sensitive domains such as mental health.

\subsection{Limitations and Future Work}
This study has several limitations. First, our analysis relies on synthetic conversations, which, despite validation against the HOPE dataset, exhibit systematic differences in attribute distributions and emotional dynamics. Moreover, HOPE itself is an imperfect reference dataset as it contains publicly available counseling transcripts varying in quality and structure and does not provide ground-truth personal profiles. Consequently, our validation establishes similarity to this particular corpus rather than demonstrating that the synthetic conversations are representative of real-world mental health interactions more broadly. Future work should validate synthetic conversations against additional real-world datasets, particularly datasets that provide richer contextual information and, where ethically and practically feasible, ground-truth personal attributes.

Second, we evaluate privacy risk over a fixed set of 21 attributes. While these attributes capture a broad range of demographic, socioeconomic, and health-related information, they do not exhaust the space of potentially inferable traits. Future work could expand the attribute space and investigate whether models can reliably infer more nuanced or higher-order characteristics.

Third, our inference pipeline uses a single model (\texttt{gpt-5.4}) under a specific ranking-based evaluation protocol. Although this provides a realistic and reproducible threat model, exposure estimates may differ under weaker or stronger inference models or alternative evaluation strategies. Comparing multiple inference models would help characterize how robust the observed exposure is to changes in adversarial capability.

In addition, our study focuses on exposure within individual conversations and does not consider longitudinal interactions, in which information may accumulate across sessions. Future work should examine whether repeated interactions produce substantially greater profile reconstruction than isolated conversations. Finally, our threat model centers on passive inference and does not account for active attacks, such as targeted prompting or deliberate information extraction, which may further increase exposure.

Future work should therefore extend this framework to multi-session settings, alternative inference models, and active threat models, while also experimentally evaluating the interventions proposed above. In particular, user studies could test whether inference summaries, contextual sensitivity cues, and inference-minimizing controls improve privacy understanding and agency without undermining the usefulness, trust, or accessibility of mental health support systems.

\section{Conclusion}
This work presents a systematic analysis of personal information exposure in LLM-mediated mental health conversations. By combining controlled synthetic dialogue generation with attribute inference, we quantify both what personal information is disclosed and what can subsequently be inferred from a single conversation. Our findings show that privacy exposure is substantial and pervasive. Conversations expose a large fraction of users' personal profiles, with sensitive mental health attributes exposed across nearly all conversations. This risk is also broadly distributed across levels of psychological distress, suggesting that substantial exposure is a routine risk of LLM-mediated mental health interactions rather than one limited to particularly vulnerable users.

Our findings broaden the scope of LLM privacy beyond memorization and explicit data extraction to inference over user-generated content. Privacy exposure can emerge from ordinary conversational interaction, without adversarial prompting or access to model training data. Methodologically, our framework demonstrates how synthetic conversations paired with ground-truth user profiles can enable controlled measurement of privacy exposure in sensitive domains where real-world data and ground truth are difficult to obtain. Together, these findings motivate privacy research and HCI design that account not only for what users disclose, but also for what their conversations make inferable, and that give users greater awareness and control over these inferential risks.

\bibliographystyle{ACM-Reference-Format}
\bibliography{ref}

\end{document}